\documentstyle[preprint,aps,floats,eqsecnum,epsf]{revtex}
\tightenlines 

\def\bp{{\bf p}}

\let\al=\alpha
\let\be=\beta
\let\ga=\gamma

\let\de=\delta
\let\De=\Delta
\let\ep=\varepsilon

\let\th=\theta

\let\<=\langle
\let\>=\rangle

\def\dsp{\displaystyle}

\let\ad=\dagger
\def\e{ {\rm e} }
\def\beq{\begin{equation}}
\def\eeq{\end{equation}}
\def\ba{\begin{array}}
\def\bea{\begin{eqnarray}}
\def\ea{\end{array}}
\def\eea{\end{eqnarray}}

\def\comment#1{ \hbox{[{\it Comment suppressed here.}\/]} }
\def\hide#1{}

\newsavebox{\eqlabel}

\makeatletter  
\newlength{\numblen}
\newsavebox{\eqnumb}
\def\@eqnnum{\savebox{\eqnumb}{\rm (\theequation)}%
\settowidth{\numblen}{\usebox{\eqnumb}}%
\makebox[\numblen][l]{\usebox{\eqnumb}~~~\usebox{\eqlabel}}}
\makeatother   

\newenvironment{equationwithlabel}[1]{ %
  \savebox{\eqlabel}{#1}
  \begin{equation}\label{#1} }{\end{equation}} 
\newcommand{\beql}[1]{\begin{equationwithlabel}{#1}}
\newcommand{\eeql}{\end{equationwithlabel}}

\begin{document}

\title{QCD at Finite Baryon Density: Nucleon Droplets and 
Color Superconductivity}

\author{Mark Alford$^1$, Krishna Rajagopal$^2$ and Frank Wilczek$^1$}
\address{~\\$^1$Institute for Advanced Study, 
        School of Natural Science \\ 
        Olden Lane, Princeton, NJ 08540}
\address{~\\$^2$Center for Theoretical Physics,\\
	Laboratory for Nuclear Science and Department of Physics\\
	Massachusetts Institute of Technology, Cambridge, MA 02139}


\maketitle

\begin{abstract}
We use a variational procedure to study finite density QCD
in an approximation in which the interaction
between quarks is modelled by that induced by instantons.
We find that uniform states with
conventional chiral symmetry breaking have negative
pressure
with respect to empty space at all but the lowest
densities, and are therefore unstable.
This is 
a precisely defined phenomenon which motivates 
the basic picture of hadrons assumed in the MIT bag model, with
nucleons as droplets of chiral symmetry restored phase.  
At all densities high enough that the chirally symmetric phase
fills space, we find that color symmetry is broken
by the formation of a $\langle qq \rangle$ 
condensate of quark Cooper pairs.
A plausible ordering scheme leads to a
substantial gap in a Lorentz scalar channel involving
quarks of two colors, and a much smaller gap in 
an axial vector channel involving quarks of the third color.
\end{abstract}
\setcounter{page}{0}
\thispagestyle{empty}
\vfill
\noindent IASSNS-HEP-97/119 

\noindent MIT-CTP-2695
\vfill
\eject

\section{Introduction}

The behavior of QCD at high density is of fundamental interest
and has potential applications to cosmology, to the astrophysics of
neutron stars, and to heavy ion collisions.  One can
make a heuristic, but we think extremely plausible, case that
essentially new forms of ordering will emerge in this regime.  

Motivated by asymptotic freedom, let us suppose as a starting point  
that at high density quarks behave nearly freely and form large Fermi 
surfaces\cite{collins}.
As we
turn on the interactions, we notice that most of the important interquark
scattering processes  
allowed by the conservation laws and Fermi
statistics 
involve large momentum transfer and are therefore
weak at asymptotically high density.  
No matter how weak the attraction, pairing of the
BCS type can be expected if there is an attractive channel\cite{BCS}.
Pairs of quarks cannot be color singlets,
and so a $\langle qq \rangle$ condensate inevitably breaks
color symmetry.  This breaking is analogous to the breaking of
electromagnetic gauge invariance in superconductivity, and might be
called
`color superconductivity'.  In this phase, the Higgs mechanism
operates and (some) gluons become massive. 
The proposed symmetry breaking in diquark channels is of course
quite different from chiral symmetry breaking
in QCD at zero density, 
which occurs in color singlet quark-antiquark channels.  

Our goal here is to explore such new forms of ordering in a context
that is definite, qualitatively reasonable, and yet sufficiently
tractable that likely patterns of symmetry breaking and rough
magnitudes of their effects can be identified concretely.  In the
course of looking for new patterns we will need to discuss the fate of
the old one, and here a surprise emerges: we find that the uniform
phase with broken chiral symmetry is unstable at any nonzero density.
At all but the lowest densities, this instability is signalled by
negative pressure, which presumably triggers the break-up of the
uniform state into regions of high density separated by empty space.

\section{Model}

Several of the methods that are used to good effect, either
quantitatively or qualitatively,  in analyzing nonperturbative QCD at
zero density,  do not seem  well adapted to finite density.
Lattice gauge theory simulations have given very limited results,
fundamentally because all known methods having acceptable efficiency
rely on importance sampling by Monte Carlo techniques which
require positivity of the action configuration by
configuration, and this positivity fails at nonzero (real) chemical
potential.  Extrapolation from supersymmetric models seems hopeless,
simply because the nuclear world would be a very different place if
one had bosonic quarks (or even baryons). 
Large $N_c$ methods are suspect, first  because one expects even-odd
effects depending on the baryon statistics and furthermore because 
we shall see concrete effects that depend critically on the
precise value $N_c = 3$.

We already briefly alluded to the fact that asymptotic freedom
suggests qualitatively new types of order at very high density.  
This regime has been studied
\cite{bailin}
by approximating the interquark
interactions by one gluon exchange, which is in fact attractive
in the color antitriplet channel.
Perturbative treatments cannot, by their nature,
do full justice to a problem
whose main physical interest arises at moderate
densities.  To get more insight into the phenomena, and in particular
to make quantitative estimates, it seems appropriate 
to analyze a tractable, physically motivated model.

In this letter we present results obtained from a variational
treatment of a
two-parameter class of models
having two flavors and three colors of massless quarks.  
We leave details of the calculations to a longer paper in
preparation. The kinetic part of the Hamiltonian is 
that for free quarks, while the
interaction Hamiltonian is a slight idealization
of the  
instanton vertex\cite{thooft} from QCD, explicitly:
\beq
H_I = -\dsp K \int\!d^3x\, 
 \bar\psi_{R1\al}\,{\psi_{Lk}}^\ga \,
 \bar\psi_{R2\be}\,{\psi_{Ll}}^\de \, \ep^{kl} 
 \, (3 {\de^\al}_\ga {\de^\be}_\de - 
 {\de^\al}_\de {\de^\be}_\ga)\ + \ {\rm h.c.}\ ,
\eeq
where $1,2,k,l$ are flavor indices,
$\al,\be,\ga,\de$ are color indices,
repeated indices are summed, and the spinor indices are
not shown. 
The overall sign is chosen negative for later convenience,
so that $K>0$ results in chiral symmetry breaking.
$H_I$ is not yet a good representation of the instanton
interaction in QCD: 
in order to mimic the effects of asymptotic freedom,
we must modify it in such a way that the interaction
decreases with increasing momentum.  We 
write $H_I$ as a mode expansion in momentum
space involving creation and
annihilation operators and spinors, and  
multiply the result by a product of form factors each of the form
\begin{equation}
F(p) ~=~ \Biggl( {\Lambda^2\over p^2 + \Lambda^2} \Biggr)^\nu\ ,
\end{equation}
one for each of the momenta of the four fermions.
This
factorized form is taken for later convenience,
and is an idealization.   $\Lambda$, of
course, is some effective QCD cutoff scale, which one might anticipate
should be in the range 300 - 1000 MeV.  $\nu$ parametrizes the shape
of the form factor; we consider $\nu=1/2$ and $\nu=1$.
Since the  
interaction we have chosen is not necessarily an accurate rendering of QCD,
\footnote{
For example, one could use a four-fermion interaction 
based on one gluon exchange \cite{iwaiwa}
} 
we will have faith only in conclusions that are robust with respect to
the parameter choices.  

The color, flavor, and Lorentz structure of our interaction has been
taken over
directly from the instanton vertex for two-flavor QCD.  
For our purposes it is very important that this interaction properly
reflect the chiral symmetry of QCD:
axial baryon number symmetry is broken, while chiral $SU(2)\times
SU(2)$ is respected.  Color (but {\it not} axial color)
is realized as a global symmetry.
There are other four-fermion
interactions in addition to $H_I$ which respect the unbroken symmetries
of QCD; 
using $H_I$ alone is the simplest way of breaking all
symmetries broken by QCD, and is therefore a good starting point. 
This model of fermions interacting via a four-fermion interaction
is
one in the long line of such
models inspired by the work of BCS as adapted to particle physics,
starting with Vaks and Larkin\cite{vaks} and 
Nambu and Jona-Lasinio\cite{njl} and studied subsequently by
many others\cite{klevansky}.  
There is also a tradition,
going
back more than twenty years and flourishing now perhaps as never
before, to model the low-energy dynamics of QCD, and specifically the
dynamics of chiral symmetry breaking, with instanton interactions
among quarks derived semi-microscopically\cite{instantonliquid}.  
This approach
cannot explain the other main qualitative aspect of low-energy QCD
dynamics, that is strict confinement of quarks, but it is
adequate for many purposes and is generating
an impressive phenomenology both of real and of numerical
experiments.

\section{Chiral Symmetry Breaking and Restoration}

We will first consider symmetry breaking in the familiar pattern known 
for QCD at zero density.  Working first at zero density, 
we choose a  variational wave function of
the form
\begin{equation}
\ba{rl}
|\psi\rangle = 
\dsp\prod_{\bp,i,\alpha}
& \Bigl( \cos(\th^L(\bp)) + \e^{i\xi^L(\bp)}\sin(\th^L(\bp))
 a^\ad_{L\,i\alpha}(\bp)  b^\ad_{R\,i\alpha}(-\bp) \Bigr) \\[1ex]
& \Bigl( \cos(\th^R(\bp)) + \e^{i\xi^R(\bp)}\sin(\th^R(\bp))
 a^\ad_{R\,i\alpha}(\bp)  b^\ad_{L\,i\alpha}(-\bp) \Bigr)|0\rangle \ ,
\ea
\label{varwf}
\end{equation}
with variational parameters $\theta$ and $\xi$ depending on the
modes.  $a^\dagger_{i\alpha}$ and $b^\dagger_{i\alpha}$ create particles
and antiparticles respectively, with flavor $i$ and color $\alpha$. 
This standard pairing form preserves the normalization of the
wave function.  The pairing occurs between particles and antiparticles with
the same flavor and color but opposite helicity and  
opposite 3-momentum.
It is in a Lorentz scalar, isospin
singlet component of the chiral $(2,2)$ representation, {\it i.e}. has
the $\sigma$-field quantum numbers standard in this context.
Following a well-trodden path we
find that the energy is minimized when
\beq
\tan(2\th^L(\bp))=\tan(2\th^R(\bp))=\frac{F^2(p) \Delta}{p}\ ;\ \ \ \ 
\xi^L(\bp) +  \xi^R(\bp) = \pi \ .
\label{variation}
\eeq
The full specification of $\xi^R$ and $\xi^L$ depends on
spinor conventions.
The gap parameter $\Delta$ is momentum independent and
is defined by
\beq
\Delta \equiv 16 K \int_0^\infty \frac{p^2 dp}{2\pi^2} F(p)
\sin(\theta(p))\cos(\theta(p))\ .
\label{deltadef}
\eeq
(\ref{variation}) and (\ref{deltadef}) are consistent only
if $\Delta=0$ or if $\Delta$ satisfies the gap equation
\beq
1 = 8K \int_0^\infty \frac{p^2 dp}{2 \pi^2} 
{ F^4(p) \over \sqrt{ F^4(p)\De^2 + p^2}},
\label{gapeq}
\eeq
Note that, unlike for standard pairing at a Fermi surface, this
equation does not have solutions for arbitrarily weak coupling, but
only for couplings above a certain threshold value.  
Also note that
although it is common to refer to $\Delta$ as the gap parameter, 
it is best thought of as inducing an effective quark mass,
which takes the form $\Delta F^2(p)$.  In the
interests of simplicity of presentation, we quote
all results in this paper for $\Delta = 400$ MeV;
we have verified that the picture we present is
qualitatively unchanged for $300$ MeV $<\Delta <$ $500$ MeV.
Fixing $\Delta$ 
fixes the magnitude (and sign) of the 
coupling $K$ for each $\Lambda$ and $\nu$.  

We now 
generalize this calculation to nonzero quark number density $n$.
It is most favorable energetically to fill 
the particle states up
to some Fermi momentum $p_F$, while leaving the
corresponding antiparticle states empty.  
That is, we replace $|0\>$ on the right hand side of (\ref{varwf})
by $|p_F\rangle$, in which all particle states with $|\bp|<p_F$
are occupied.  
The quark number density, 
\beq
n=\frac{2}{\pi^2}\, p_F^3\ ,
\label{ndef}
\eeq
is determined by $p_F$
but as we shall see $p_F$ is not equal to the chemical
potential $\mu$.
The creation operators in (\ref{varwf}) 
for modes below the Fermi surface annihilate $|p_F\>$,
and so for these modes effectively $\theta(\bp)=0$.
On the other hand for
the unoccupied states, with $|\bp| > p_F$, the variational scheme is
unmodified.  Note that the condensate does not
affect $n$.
Thus we arrive at a very simple modification
of the gap equation: the lower limit of the integral in
(\ref{gapeq})  is $p_F$,
rather than 0.  Since the interaction is ever more effectively 
quenched as $p_F$ increases, the gap
parameter
$\Delta (p_F)$ arising as the solution of (\ref{gapeq}) 
will shrink monotonically
and eventually vanish as $p_F$ increases.  Let us define
$n_c$ to be the critical
density at which $\Delta$ vanishes.

Having obtained a definite wave function, we can evaluate the
energy density in terms of the gap parameter.  Relative to
the energy of the naive vacuum 
state $|0\>$, the energy density $\varepsilon(n)$ is
given by
\beq
{1\over 24}\varepsilon(n) = \frac{p_F^4}{16 \pi^2}+
\int_{p_F}^\infty \frac{p^2 dp}{2 \pi^2}  \frac{p}{2} \left(
 1 - { p \over \sqrt{\De^2 F^4(p) + p^2}} \right)
 - {\De^2 \over 32 K}\ ,
\label{chiralen}
\eeq
where $p_F(n)$ is to be obtained from (\ref{ndef}).  
The first and second terms are the kinetic energies of the
modes respectively below and above the nominal Fermi surface,
while the third term is the interaction energy.  It is startling, 
perhaps,
that in the first term the bare energy occurs, with no effective mass,
but this is a direct reflection of the situation discussed in the
previous paragraph.  For a proper interpretation, however, it is
important to note that adding a particle at the Fermi
surface modifies both the kinetic and the interaction term, and that
these conspire to give a gap or effective mass $F(p_F)^2\Delta$ for
the physical excitations there.  
The chemical
potential $\mu$, the minimum energy required to add one more quark
to the state, is given by $\mu=\partial \varepsilon/\partial n$ at fixed 
volume. We have verified that $\mu^2= p_F^2 + \Delta^2 F^4(p_F)$.
Note that $\varepsilon(0)<0$, reflecting the fact that the
physical vacuum state with its chiral condensate has a lower energy
than the state $|0\>$.  Measured relative to
the physical vacuum, the energy density at nonzero $n$ is
$e(n)\equiv \varepsilon(n)-\varepsilon(0)$.

\begin{figure}[t]
\centerline{
\epsfysize=3in
\hfill\epsfbox{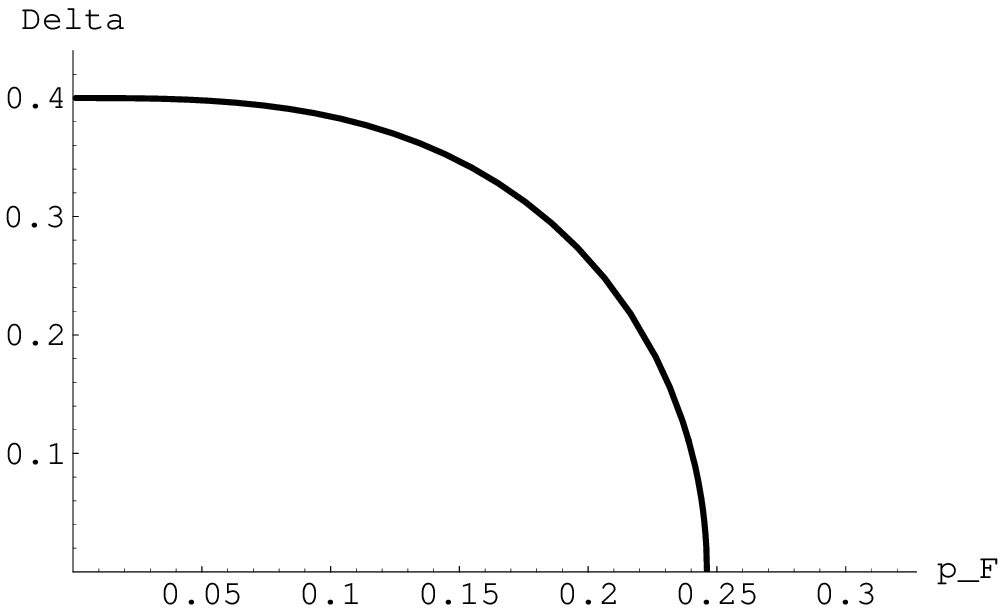}\hfill
\epsfysize=3in
\hfill\epsfbox{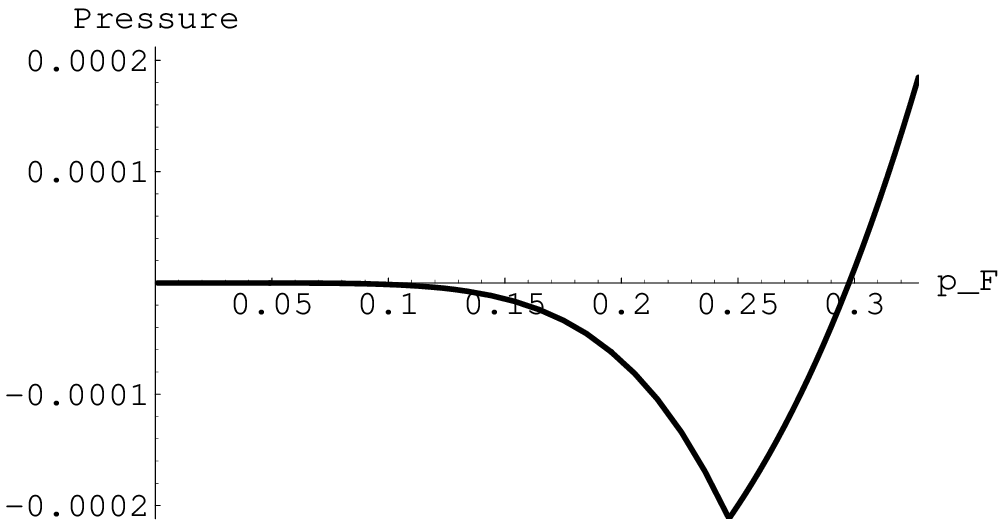}\hfill
}
\caption{Chiral gap (in GeV) and the pressure 
(in GeV$^4$) as a function
of $p_F = (n \pi^2/2)^{1/3}$ in GeV.  
The  
pressure is positive at {\it very}
small $p_F$; it becomes negative at a
$p_F$ which is less than 0.001.
At $n=n_c$,
the gap vanishes and the pressure,
which is still negative,
has a cusp. At $n=n_0>n_c$, the pressure
crosses zero and becomes positive.} 
\end{figure}
The equation for $\Delta (p_F)$ can be solved numerically, and the
resulting energy evaluated.  Great physical interest attaches to the 
pressure 
\begin{equation}
P(n) ~=~ n \, {\partial e \over \partial n}  - e =
n^2 \frac{\partial  }
 {\partial n} \left( {e\over n} \right)
\end{equation}
of a region with density $n$, where the pressure of the
physical vacuum is by definition zero.
For all values of the parameters that we consider reasonable,
we find that after a tiny interval of very low densities
at which the pressure is positive, {\em the pressure  
becomes negative, and continues to decrease until the
critical density $n_c$ 
at which chiral symmetry is restored}.  At that point we
switch over to an essentially free quark phase, in which
the energy density relative to the physical vacuum
is $3 p_F^4 / (2\pi^2)\, - \,\varepsilon(0)$,
and the pressure is given by $p_F^4/(2\pi^2)\, +\, \varepsilon(0)$. 
At $n=n_c$, where these phases join,
the pressure function is continuous, with a finite
negative value, but it has a cusp.  
As density increases in the
free-quark phase, the pressure increases monotonically, and
passes through zero again at some density $n_0 > n_c$.  
At $n=n_0$, the energy per quark $e/n$ is minimized  and this phase
is stable at zero pressure.
Representative
results are recorded in Table 1, and $\Delta$ and $P$ are shown 
for one set of parameters in Figure 1. 
\def\st{\rule[-1.5ex]{0em}{4ex}} 
\begin{table}[t]
\begin{tabular}{ccccc}
\st$\nu$ & $\Lambda$ & $p_F(n_c)$ & $p_F(n_0)$ & $\varepsilon(0)$  \\[1ex]
\hline
 1 & 0.6 & 0.212 & 0.268 & $-(0.127)^4$ \\
 1 & 0.8 & 0.246 & 0.298 & $-(0.141)^4$ \\
 1 & 1.0 & 0.274 & 0.321 & $-(0.152)^4$ \\
\hline
 0.5 & 0.6 & 0.262 & 0.310 & $-(0.147)^4$ \\
 0.5 & 0.8 & 0.298 & 0.340 & $-(0.161)^4$ \\
 0.5 & 1.0 & 0.328 & 0.362 & $-(0.172)^4$ \\
\end{tabular}
\vspace{3ex}
\caption{Comparison of the Fermi momenta corresponding to
the density $n_c$ at which the chiral gap vanishes,
the density $n_0$ at which there is a stable zero pressure phase,
and the energy density $\varepsilon(0)$ of the vacuum phase, 
for six choices of parameters. 
In each case the coupling is fixed by requiring $\Delta= 0.4$ GeV at $n=0$.
All numbers are in GeV.
}
\label{tableofresults}
\end{table}


Evidently, at all but the lowest densities (which
we discuss below) in the presence of
a chiral condensate the negative pressure
associated with increasing vacuum energy
overcompensates the increasing Fermi pressure. 
This negative pressure
signals mechanical instability of the uniform
chiral symmetry broken phase.    
There is
an attractive physical interpretation of this phenomenon.  
The
uniform nonzero density phase will break up into 
stable droplets of high
density $n=n_0$ in which the pressure is zero 
and chiral symmetry is restored, surrounded by  
empty space with $n=P=0$.  
There are preliminary indications of this behavior in numerical
simulations of a model similar to ours\cite{aichelin}.
Although our simple calculations do not
allow us to
follow the evolution and eventual stabilization of the original quark
cloud, it is hard  
to avoid identifying the droplets of chiral
symmetric phase into which it condenses with physical 
nucleons.  Nothing within the model tells us that
the stable droplets have quark number $3$; nucleons are simply
the only candidates in nature which can be identified with droplets
within which the quark density is nonzero and the chiral condensate
is zero. 
If correct, this identification is very reminiscent of the MIT
bag philosophy, 
here arising in the description of a sharply defined physical
phenomenon.\footnote{Considerations similar
to those we describe also lead Buballa\cite{buballa}
to conclude that 
in a Nambu Jona-Lasinio model
with an interaction which differs from the one we use,
matter with broken chiral symmetry is unstable
and nucleons 
can therefore only be viewed as bags within which chiral symmetry
is restored.}
It seems quite different, at least superficially, from
the Skyrme model, where the chiral symmetry order parameter changes in
direction but not in magnitude within the nucleon.

What, then, of the positive
pressure phase at very low density?  Without external pressure,
this dilute gas of quarks with mass $\Delta(0)$
would expand and dissipate. Even if some external pressure prevents
expansion, however, this phase is only metastable: 
its energy per quark $e/n \sim \Delta(0)$ is greater
than that in the stable phase at $n=n_0$, which 
satisfies
\beq
\frac{e(n_0)}{n_0}=p_F(n_0)
\label{eovern}
\eeq
and can therefore be read off Table I.
When fluctuations in this dilute gas 
increase the density in a region enough that the pressure
becomes negative, this region can collapse to 
density $n_0$. In this way, the metastable phase
converts to regions of the stable phase at $n=n_0$,
surrounded by vacuum.
(Note that at the density at which the pressure first
becomes negative, $e/n$ is at a local maximum and so
this zero pressure phase is unstable, unlike that at
$n=n_0$.)
We see that the fate of the low density positive pressure phase
is the same as that of the negative pressure phase.
{\it Any} uniform phase with chiral symmetry broken 
evolves into an inhomogenous mixture of droplets
within which $n=n_0$ and chiral symmetry is
restored, surrounded by regions of
vacuum. 

The satisfying picture just discussed is not
obtained for all parameter values, however.
For example, for $\nu=1$ and $2.2 < \Lambda < 3.2$  
the phase at $n=n_0$ has higher $e/n$ than
that of a  dilute gas of quarks with mass
$\Delta(0)$.  For $\Lambda > 3.2$ GeV,
the pressure is positive for all $n$.  In these 
(fortunately, unreasonable) parameter ranges,
the model, without further
modification,  has no reasonable physical interpretation.
 
At a quantitative level, a naive implementation of our proposed
identification of droplets of $n=n_0$ matter with nucleons
works surprisingly well.   
One
might want to identify $n_0$ with the quark density at the
center of  baryons.  Taking this to be three times nuclear matter
density yields $p_F(n_0)\sim 0.39 {\rm ~GeV}$. 
On the other hand, requiring $e(n_0)/n_0$ to be one third
the nucleon mass yields $p_F(n_0)\sim 0.31$ GeV.
Our toy model treatment cannot meet both criteria simultaneously, which
is not surprising,
but we see from Table I that the magnitude of $p_F(n_0)$
is very reasonable.
The 
vacuum energy, which becomes the bag constant,
is also of the correct order of magnitude.
Adding further interactions to $H_I$
would obviously
make a quantitative difference, 
but there is no
reason to expect the qualitative picture to change.

In any case, the physical picture suggested here has significant
implications for the phase transition, as a function of density, to
restored chiral symmetry.   Since the nucleons are regions where the
symmetry is already restored, the transition should occur by
a mechanism analogous to percolation
as nucleons merge\cite{satz}. This transition should
be complete once a density characteristic of the center of 
nucleons is achieved.  
The fact that some external pressure must
be imposed in order to induce the nucleons to merge (e.g. the
fact that in nuclear matter at zero pressure the 
nucleon droplets remain unmerged)
must reflect interactions between droplets, which we have not treated here.
The mechanism of chiral symmetry restoration at finite density but
zero temperature is quite different from the one we expect at finite
temperature and zero density: it occurs by percolation 
among pre-formed bags of symmetric phase.  
Of course, this is no contradiction,
because the finite temperature transition occurs at such a low
temperature that few baryons are present.

\section{Color Superconductivity}

At high density, pairing of particles near the Fermi surface as in the
original BCS scheme\cite{BCS} becomes more favorable.  
Our Hamiltonian supports
condensation in quark-quark channels.   
The condensation is now between identical fermions with the same
helicity, and the Hamiltonian selects chiral isosinglets --- that is,
antisymmetry in flavor.  
One can therefore have spin 0 --- antisymmetric in spin and therefore
in color, forming a $\bar {\bf 3}$, or spin 1 --- symmetric in spin and
therefore in color, forming a {\bf 6}.   

We first consider the former.  A suitable trial wave function is
\beq
|\psi\> = G_L^\ad G_R^\ad |p_F\>
\eeq
where
\beq
\ba{rcll}
G_L^\ad &=& \dsp
 \prod_{\alpha,\beta,\bp}
& \Bigl( \cos(\th^L_{A}(\bp)) + \ep^{\alpha\beta 3}\e^{i\xi^L_A(\bp)}
\sin(\th^L_{A}(\bp))
 a^\ad_{L\,1\alpha}(\bp)  a^\ad_{L\,2\beta}(-\bp) \Bigr) \\[1ex]
&&& \Bigl( \cos(\th^R_{B}(\bp)) + \ep^{\alpha\beta 3}\e^{i\xi^R_B(\bp)}
\sin(\th^R_B(\bp))
 b^\ad_{R\,1\alpha}(\bp)  b^\ad_{R\,2\beta}(-\bp) \Bigr) \\[1ex]
&&& \Bigl( \cos(\th^R_{C}(\bp)) + \ep^{\alpha\beta 3}\e^{i\xi^R_C(\bp)}
\sin(\th^R_{C}(\bp))
 a_{R\,1\alpha}(\bp)  a_{R\,2\beta}(-\bp) \Bigr) \\[1ex]
G_R^\ad &=& \multicolumn{2}{l}{\hbox{same, with~~}R\leftrightarrow L}.
\ea
\label{colwf}
\eeq
Here,
$\alpha$ and $\beta$ are color indices, and we
have chosen to pair quarks of the first two colors.
$1$ and $2$ are flavor indices. The first term in (\ref{colwf})
creates particles above the Fermi surface; the second creates
antiparticles; the third creates holes below the Fermi surface.
In this state, the Lorentz scalar
$\langle q^{i\alpha} \,C\gamma^5 q^{j\beta}\varepsilon_{ij}\,
\varepsilon_{\alpha\beta 3}\rangle$ is nonzero. This 
singles out a preferred
direction in color space and breaks color $SU(3) \rightarrow SU(2)$.
The $U(1)$ of electromagnetism is spontaneously broken but
there is a     linear combination of electric charge
and color hypercharge under which the condensate is neutral, and
which therefore generates an unbroken
$U(1)$ gauge symmetry.  No flavor symmetries,
not even the chiral ones,  are broken.

Note that $n$ is now not given by (\ref{ndef}) because
the operators in (\ref{colwf}) can change particle number.
Varying the expectation value of $H-\mu N$ in this state
with respect to $p_F$ yields $p_F=\mu$, unlike in 
the case of the chiral condensate.  This difference
reflects the fact that 
a gap in a $\<qq\>$ channel does not act as an effective mass
term in the way that a chiral gap                  
does.  Upon adding a quark, the condensate can adjust in such
a way that the energy cost is only $p_F$.   $\Delta$
is, however, a true gap in the sense of
condensed matter physics: the energy cost of 
making a particle-hole excitation is  
$2\Delta$ at minimum.
Varying with respect to 
all the other variational parameters yields
$\xi^R_{A,B,C} + \xi^L_{A,B,C} = \pi$, $\theta^R_{A,B,C}=
\theta^L_{A,B,C}$, and
\beq
\tan(2\theta^L_A(\bp)) = \frac{F^2(p)\Delta}{p-\mu}\ ,
\ \ \
\tan(2\theta^L_B(\bp)) = \frac{F^2(p)\Delta}{p+\mu}\ ,
\ \ \
\tan(2\theta^L_C(\bp)) = \frac{F^2(p)\Delta}{\mu-p}\ .
\label{colorvariation}
\eeq
Here, the gap $\Delta$ satisfies
a self-consistency equation of the form
\beq
\ba{rrl}
1 = \dsp{2K} \biggl\{ &&
 \dsp\int_\mu^\infty \frac{p^2 dp}{2\pi^2} { F^4(p) 
\over\sqrt{ F^4(p)\De^2 + (p-\mu)^2}}\\[3ex]
& + &\dsp\int_0^\infty\frac{p^2 dp}{2\pi^2} 
{ F^4(p) \over \sqrt{ F^4(p)\De^2 + (p+\mu)^2}}\\[3ex]
& + &\dsp\int_0^\mu\frac{p^2 dp}{2\pi^2} 
{F^4(p) \over \sqrt{ F^4(p)\De^2 + (\mu-p)^2}}
\ \ \biggr\}\ .
\ea
\label{colorgapeq}
\eeq
The three terms in this equation arise respectively from 
particles above the 
Fermi surface, antiparticles, 
and particles below the Fermi surface.  
For $\mu >  0 $ the particle and hole integrals diverge logarithmically
at the Fermi surface 
as $\Delta \rightarrow 0$, which signals the possibility of
condensation for arbitrarily weak attraction.

The numerical coefficient in this color superconducting gap equation
is smaller than the corresponding coefficient in the chiral symmetry
breaking case.  The exact factor follows from the precise form of the
Hamiltonian, but part of the explanation is simple and robust: 
chiral condensation makes good use of all three colors coherently,
but the color superconducting condensation, which breaks color
symmetry, cannot.

One can form reasonable qualitative expectations for the solution of
the gap equation without detailed calculations.  
Because the numerical  coefficient in
the gap equation is smaller than
the threshold value at which one would
have a nonzero $\Delta$ at $\mu=0$, $\Delta$ would be zero were
it not for the logarithmically divergent contribution
to the integral from the region near $\mu$. 
This means that at small $\mu$, the gap must be small
because the density of
states at the Fermi surface is small.  This has only formal
significance, because the only densities of physical relevance
are $n=0$ and $n\ge n_0$. At intermediate densities, matter
is in an inhomogeneous mixture of the $n=0$ and $n=n_0$
phases. (We are assuming that the 
color breaking condensate does not significantly
affect $n_0$; this will be discussed below.)
As $\mu$ increases, the density of states at
the Fermi surface increases and the gap parameter grows.  Finally, at
large $\mu$ the effect of the form factor $F$ is felt,
the effective coupling decreases, and the
gap parameter goes back down.  For the parameter ranges we have examined
the gap parameter is quite substantial:  $\sim 50-150$ 
MeV at $n_0$, and
peaking at $100-200$ MeV at a density somewhat higher. We plot
$\Delta$ for two sets of parameters in Figure 2.  The density
at which the gap peaks depends on $\Lambda$; the shape
of the curve depends on $\nu$; the height of the curve is
almost independent of both.
\begin{figure}[t]
\centerline{
\epsfysize=3in
\hfill\epsfbox{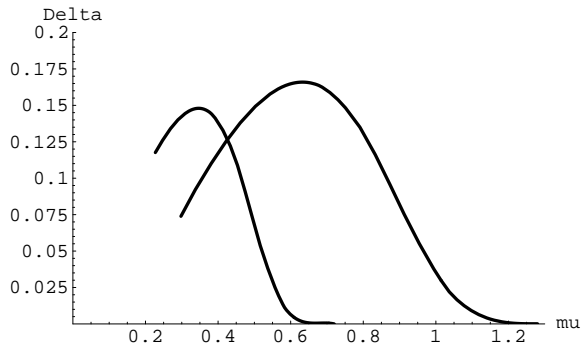}\hfill
}
\caption{Gap created
by the Lorentz scalar color superconductor 
condensate,
as a function of $\mu=p_F$ for $\nu=1$ and (from left to right) 
$\Lambda=0.4,0.8$ GeV. Each curve begins
where $n$ is given by the appropriate $n_0$.}
\end{figure}

As in the previous section, 
one can obtain expressions for the energy and the density,
and thus derive the equation of state.  We find that the equation of
state is hardly modified from the free-quark values --- the pressures,
at equal density, are 
equal to within a few per cent.  This makes it very plausible that,
as we assume, 
the color condensation makes only a small change in the 
density $n_0$ at which a stable phase exists at zero pressure. 
To make the argument rigorous,
we must do a calculation in which we consider chiral and
color condensation simultaneously; we should form 
a trial wave function that allows for both possibilities
and allows them to compete.  We have
begun this calculation, but will not report on it here
other than to note that since the two condensates compete
for the same quarks the bigger of the two tends to suppress
the smaller.  This is further evidence that the potential for
the formation of  a color breaking gap does not affect the result
that $n_0>n_c$.
For practical purposes, it appears to be a very good approximation to
treat the condensations separately, as we do here,
because where one is large the other is small, even before they compete.

The third color has so far been left out in the cold, but we can gain
energy by allowing it too to condense.  The available channel is the
color {\bf 6}.  We find that there is attraction in this channel, for a
pairing that is a spatial axial vector.  
Thus we predict that 
not only color, but also
rotation invariance, is spontaneously broken by QCD at high density.
Axial vector quantum numbers, of course, are characteristic of spin
alignment.  The appropriate trial wave function for this condensation
is obtained by acting on $|p_F\rangle$ by a product
of operators analogous to the product in (\ref{colwf})
except that they now involve
only the third color and depend on
the direction in momentum space.  The 
particle pair
creation operator, for example, can be written
\begin{equation}
\frac{1}{\sqrt{1+\Bigl[\th_L^A(|\bp|)p_z/|\bp|\Bigr]^2}} 
\Biggl(\,1\, +\, \th^L_{A}(|\bp|)\,\frac{p_z}{|\bp|}\,
\e^{i\xi^L_A(\bp)}
 a^\ad_{L\,13}(\bp)  a^\ad_{L\,23}(-\bp) \Biggr) \ .
\end{equation}
In this state, the nonzero condensate is 
$\langle q^{i3} \, C
\sigma_{03}\, q^{j3}\varepsilon_{ij}\rangle$, where
$\sigma_{\mu\nu} = (i/2)[\gamma_\mu,\gamma_\nu]$.
This is an axial vector pointing in the $z$-direction,
manifestly breaking rotation invariance. It does not change sign
under spatial inversion.
It breaks the gauged $U(1)$ symmetry left unbroken
by the scalar superconductor condensate, but does
not affect the unbroken $SU(2)$ subgroup of
color $SU(3)$.  

Finding the self-consistent solution to 
the variational problem for the axial condensate
is somewhat involved, because of the expressions which arise
when the angular integrals in $\langle H \rangle$ are performed.
The expression analogous to $\tan(2\theta)$ of (\ref{variation})
cannot explicitly be inverted, so it is not possible to
reduce the two equations analogous to (\ref{variation})
and (\ref{deltadef})  to a single gap equation.
The results are nevertheless
easily described.         
We define a momentum independent parameter $\Delta$
which can be viewed as the average of the gap over the
Fermi surface, and which satisfies self-consistency
relations. The effective coupling is smaller than
in the scalar channel for two reasons. First, only
one color participates. Second, most regions of 
the Fermi surface do not participate fully.
We 
find that the gap must be pushed to very
small values.  Because of the logarithmic singularity
in the integrals, the resulting gap is exponentially
sensitive to parameter choices.  We find 
$\Delta$ of order a few keV at most, but 
this should not be considered a robust result.  It is worth exploring
whether plausible interactions can be added to $H_I$ which
have the effect of strengthening the axial vector condensate,
particularly as such a condensate could lead to signatures
in heavy ion collisions. 
The  existence of a preferred direction for spins could be observable,
if it were reasonably efficiently handed down to $\Lambda$ baryons,
as it would lead to
correlation between the polarization of different $\Lambda$'s
in a single event.
Without modification, our model suggests
that this axial vector condensate
is {\it much} smaller than the scalar color breaking
condensate.  A gap this small is surely irrelevant
in heavy ion collisions, but has both        formal
consequences and implications for neutron star physics.
It is striking that 
a single interaction generates coexisting condensates with
scales which differ by five orders of magnitude.

In our model as it stands, color 
is realized as a global symmetry.  Breaking of this symmetry generates
Nambu-Goldstone bosons, formally.  
However,
in reality color
is of course
a gauge symmetry, and the true spectrum does not contain massless
scalars, but rather massive vectors.   Aside from a node along the
equator of the Fermi surface
for one color, there is a gap everywhere on the quark Fermi
surfaces. 
To this point, we have described the color superconducting
phase as a Higgs phase.  One expects, however, that 
there is a complementary description in which this
is a confining phase, albeit one with two vastly different
confinement lengths, neither of which is related to the
confinement length at zero density.
%
As a formal matter, it is 
of some interest that the color superconducting phase can be considered a
realization of {\it confinement without chiral symmetry breaking}. 

In looking for signatures of color superconductivity in heavy ion physics
and in neutron stars, it is unfortunate that the equation
of state is almost equal to that for a deconfined phase
with no diquark condensate.
Superconducting condensates do modify the gauge interactions and
this may have implications in heavy ion collisions.
The scalar condensate carries electric as well as color charge.  It
is neutral under a certain combination of electrodynamic and color
hypercharge, so taken by itself it would leave a modified massless
photon.   
If densities above $n_0$ are achieved at
low enough temperatures that the scalar condensate forms,
there will
be a mixture of the ordinary photon and the color hypercharge gauge
boson which is massless.
(This modified photon would acquire a small mass from the 
axial vector condensate if temperatures were low enough
for this condensate to be present.)  
There is also
a residual $SU(2)$ gauge symmetry,
presumably deconfined,
and there are five gluons whose mass is set by the scalar 
condensate.   
Either the modification of the photon
or the loss of massless gluons could 
have consequences, but dramatic effects
do not seem apparent.

Turning to neutron 
star phenomenology, there is some indication, from the slowness
of observed neutron star cooling rates, that a gap in the excitation
spectrum for quark matter might be welcome\cite{ogleman},  
as this suppresses neutrino emission via weak interaction
processes involving single
thermally excited $u$ and $d$ quarks by $\exp(-\Delta/T)$.
A $400$ keV gap has dramatic consequences\cite{schaab};
the scalar gap is therefore enormous in this context, and the
axial vector gap plays a role too, shutting down
these direct neutrino emission processes completely
once the core cools to temperatures at which
the axial condensate forms. It would also be
worthwhile to explore the effects of
the presence of
macroscopic regions in which an axial vector
condensate is ordered.

\section{Discussion}

Many things were ignored in this analysis.  Most important, perhaps,
is the strange quark.  In the spirit of the analysis, we should
consider the modified instanton vertex including the strange quark as
well.  This adds an incoming left-handed and an outgoing right-handed
leg.  
If the mass of this quark were large, we could connect these legs with
a large coefficient, and reduce to the previous case,
perhaps with an additional four-fermion
vertex involving all
three flavors modelled on one-gluon exchange.  
Whatever the interaction(s), color superconductivity in
a three flavor theory necessarily introduces the new
feature of flavor symmetry breaking. Both the 
condensates considered in this paper are flavor singlets;
this is impossible for a $\langle qq \rangle$ condensate
in a three flavor theory.  One particularly attractive
possibility
is condensation in the 
$\langle q^\alpha_i \, C \gamma^5 q^\beta_j \varepsilon^{ijA}
\varepsilon_{\alpha\beta A}\rangle$
channel, with summation over $A$.
This breaks 
flavor and color in a coordinated fashion,
leaving unbroken the
diagonal subgroup of $SU(3)_{\rm color}\times SU(3)_{\rm flavor}$.

Another question concerns the postulated Hamiltonian.  While there are
good reasons to take an effective interaction of the instanton type
as a starting point, there could well be 
significant corrections affecting the more delicate
consequences such as axial vector {\bf 6} condensation.   A specific,
important example is to compare the effective interaction derived from
one-gluon exchange.   It turns out that this interaction has a
similar pattern, for our purposes, to the instanton: it is very attractive in
the $\sigma$ channel, attractive in the color antitriplet scalar, and
neither attractive nor repulsive in the color sextet axial vector.
More generally, it would be desirable to use a renormalization
group treatment to find the interactions which are
most relevant near the Fermi surface.

The qualitative model we have treated suggests a compelling picture
both for the chiral restoration transition and for the color
superconductivity which sets in 
at densities just beyond. It
points toward future
work in many directions:  the percolation transition
must be characterized; consequences in neutron 
star and heavy ion physics remain to be elucidated;
the superconducting ordering patterns
may hold further surprises, particularly
as flavor becomes important. 
The whole subject needs more work; the microscopic
phenomenon is so remarkable, that we suspect our imaginations have
failed adequately to grasp its implications. 

\acknowledgments

Related work on color superconductivity 
has been done independently by R. Rapp,
T. Schaefer, E. V. Shuryak and M. Velkovsky. We
wish to thank them and M. Gyulassy, R. L. Jaffe, M. Prakash, 
H. Satz, M. Stephanov
and I. Zahed for helpful discussions. We thank
the RIKEN-BNL Center, where this work was completed,
for its hospitality.

The research of MA and FW
is supported in part by DOE grant DE-FG02-90ER40542;
MA is also supported by the generosity of Frank and Peggy Taplin;
that of KR is supported in part by DOE cooperative
research agreement DE-FC02-94ER40818.

\end{document}